\documentclass[fleqn,twoside]{article}
\usepackage[headings]{espcrc2}

\readRCS $Id: espcrc2.tex,v 1.2 2004/02/24 11:22:11 spepping Exp $
\usepackage{graphicx}

\newcommand{\AmS}{{\protect\the\textfont2
  A\kern-.1667em\lower.5ex\hbox{M}\kern-.125emS}}

\def\refjl#1#2#3#4#5#6{\bibitem{#1} #2, {#3} {#4} (#5) #6.}

\def\etal{{et al.}}
\def\NP{Nucl. Phys.}
\def\NPPS{Nucl. Phys. B (Proc. Suppl.)}
\def\PL{Phys. Lett.}
\def\PRL{Phys. Rev. Lett.}
\def\PR{Phys. Rev.}

\def\ZP{Z. Phys.}

\def\MP{Int. J. Mod. Phys.}

\def\RMP{Rev. Mod. Phys.}
\def\RPP{Rep. Prog. Phys.}

\def\PPNP{Prog. Part. Nucl. Phys.}

\def\EPJ{Eur. Phys. J.}
\newcommand{\eqn}[1]{(\ref{#1})}
\newcommand{\be}{\begin{equation}}
\newcommand{\ee}{\end{equation}}
\newcommand{\no}{\nonumber}
\newcommand{\bel}[1]{\be\label{#1}}
\newcommand{\ba}{\begin{array}{c}}
\newcommand{\bat}{\begin{array}{cc}}
\newcommand{\ea}{\end{array}}
\newcommand{\beqn}{\begin{eqnarray}}
\newcommand{\eeqn}{\end{eqnarray}}

\newcommand{\bi}{\begin{itemize}}
\newcommand{\ei}{\end{itemize}}

\newcommand{\gsim}{~{}_{\textstyle\sim}^{\textstyle >}~}
\newcommand{\lsim}{~{}_{\textstyle\sim}^{\textstyle <}~}

\newcommand{\cL}{{\cal L}}

\newcommand{\Br}{\mathrm{Br}}

\title{Tau Physics 2006: Summary \& Outlook}

\author{A. Pich\address{
         IFIC, Universitat de Val\`encia -- CSIC, 
         Apt. Correus 22085, E--46071 Val\`encia, Spain}
         }

\begin{document}

\begin{abstract}
\vspace{1pc}
 A large amount of new results have been presented at
TAU2006. The highlights of the workshop, the present status of a few
selected topics on lepton physics (universality, QCD tests, $V_{us}$
determination from $\tau$ decay, $g-2$, $\nu$ oscillations,
lepton-flavour violation) and the prospects for future improvements
are briefly summarized.
\vspace{1pc}
\end{abstract}

\maketitle

\section{INTRODUCTION}

The known leptons provide clean probes to perform very precise tests
of the Standard Model and search for signals of new dynamics. The
electroweak gauge structure has been successfully tested at the
0.1\% to 1\% level, confirming the Standard Model framework
\cite{SanFeliu04}.
Moreover, the hadronic $\tau$ decays turn out to be a beautiful
laboratory for studying strong interaction effects at low energies
\cite{taurev98,taurev03,Stahl00}. 
Accurate determinations of the QCD coupling, $|V_{us}|$ and the
strange quark mass have been obtained with $\tau$ decay data.

The first hints of new physics beyond the Standard Model have also
emerged from the lepton sector. Convincing evidence of neutrino
oscillations has been obtained by SNO \cite{SNO} and
Super-Kamiokande \cite{SKsolar,SKatm}. Combined with data from other
neutrino experiments \cite{KamLAND,K2K,MINOS}, it shows that
$\nu_e\to\nu_{\mu}$ and $\nu_\mu\to\nu_\tau$   
transitions do occur.

The huge statistics accumulated at the B Factories allow to explore
lepton-flavour-violating $\tau$ decay modes with increased
sensitivities beyond $10^{-7}$, which could be further pushed down
to few $10^{-9}$ at future facilities. Moreover, BESIII will soon
start taking data at the new Beijing Tau-Charm Factory. With the
excellent experimental conditions of the threshold region,
complementary information on the $\tau$ should be obtained, such as
an improved mass measurement.

The large amount of new results presented at this workshop shows
that $\tau$ physics is entering a new era, full of interesting
possibilities and with a high potential for new discoveries.

\section{LEPTON UNIVERSALITY}
\label{sec:universality}

\begin{table}[t] 
\centering \caption{Present constraints on $|g_l/g_{l'}|$
\cite{taurev03,PDG,LEPEWWG,Fiorini}.} \label{tab:ccuniv}
\vspace{0.2cm}
\renewcommand{\tabcolsep}{1.1pc} 
\renewcommand{\arraystretch}{1.15} 
\begin{tabular}{lc}
\hline & $|g_\mu/g_e|$ \\ \hline
$B_{\tau\to\mu}/B_{\tau\to e}$ & $1.0000\pm 0.0020$ \\
$B_{\pi\to \mu}/B_{\pi\to e}$ & $1.0017\pm 0.0015$ \\
$B_{K\to \mu}/B_{K\to e}$ & $1.012\pm 0.009$ \\
$B_{W\to\mu}/B_{W\to e}$  & $0.997\pm 0.010$ \\
\hline\hline & $|g_\tau/g_\mu|$  \\ \hline
$B_{\tau\to e}\,\tau_\mu/\tau_\tau$ & $1.0004\pm 0.0022$ \\
$\Gamma_{\tau\to\pi}/\Gamma_{\pi\to\mu}$ &  $0.996\pm 0.005$ \\
$\Gamma_{\tau\to K}/\Gamma_{K\to\mu}$ & $0.979\pm 0.017$ \\
$B_{W\to\tau}/B_{W\to\mu}$  & $1.039\pm 0.013$
\\ \hline\hline
& $|g_\tau/g_e|$  \\ \hline
$B_{\tau\to\mu}\,\tau_\mu/\tau_\tau$ & $1.0004\pm 0.0023$ \\
$B_{W\to\tau}/B_{W\to e}$  & $1.036\pm 0.014$
\\ \hline
\end{tabular}
\end{table}
%

In the Standard Model all lepton doublets have identical couplings
to the $W$ boson. Comparing the measured decay widths of leptonic or
semileptonic decays which only differ in the lepton flavour, one can
test experimentally that the $W$ interaction is indeed the same,
i.e. that \ $g_e = g_\mu = g_\tau \equiv g\, $. As shown in
Table~\ref{tab:ccuniv}, the present data verify the universality of
the leptonic charged-current couplings to the 0.2\% level.\footnote{
$\Br(W\to\nu_\tau\tau)$ is $2.1\,\sigma/2.7\,\sigma$ larger than
$B(W\to \nu_e e / \nu_\mu\mu)$. The stringent limits on
$|g_\tau/g_{e,\mu}|$ from $W$-mediated decays make unlikely that
this is a real effect.}
%

The $\tau$ leptonic branching fractions and the $\tau$ lifetime are
known with a precision of $0.3\%$. A slightly improved lifetime
measurement could be expected from BABAR and BELLE \cite{Lusiani}.
For comparison, the $\mu$ lifetime is known with an accuracy of
$10^{-5}$, which should be further improved to $10^{-6}$ by the
MuLan experiment at PSI \cite{LY06}.

The universality tests require also a good determination of
$m_\tau^5$, which is only known to the $0.08\%$ level. Two new
preliminary measurements of the $\tau$ mass have been presented at
this workshop:
$$
m_\tau =\left\{ \begin{array}{lr} 1776.71\pm 0.13\pm
0.35~\mathrm{MeV}
&\; [\mathrm{BELLE}],\\
1776.80\, {}^{+\, 0.25}_{-\, 0.23} \pm 0.15~\mathrm{MeV} &\;
[\mathrm{KEDR}]. \ea\right.
$$
BELLE \cite{Shapkin} has made a pseudomass analysis of
$\tau\to\nu_\tau 3\pi$ decays, while KEDR \cite{KEDR} measures the
$\tau^+\tau^-$ threshold production, taking advantage of a precise
energy calibration through the resonance depolarization method. In
both cases the achieved precision is getting close to the present
BES-dominated value, $m_\tau = 1776.99\, {}^{+\, 0.29}_{-\, 0.26}$
\cite{PDG}. KEDR aims to obtain a final accuracy of 0.15 MeV. A
precision better than 0.1 MeV should be easily achieved at BESIII
\cite{MO}, through a detailed analysis of
$\sigma(e^+e^-\to\tau^+\tau^-)$ at threshold \cite{Pedro,Voloshin}.

\section{HADRONIC TAU DECAYS}
\label{sec:hadronic}

The $\tau$ is the only known lepton massive enough to decay into
hadrons. Its semileptonic decays are then ideally suited for
studying the hadronic weak currents in very clean conditions.
The 
decay $\tau^-\to\nu_\tau H^-$ probes the matrix element of the
left--handed charged current between the vacuum and the final
hadronic state $H^-$.

\begin{figure}[tbh]
\centering
\includegraphics[angle=-90,width=7.3cm,clip]{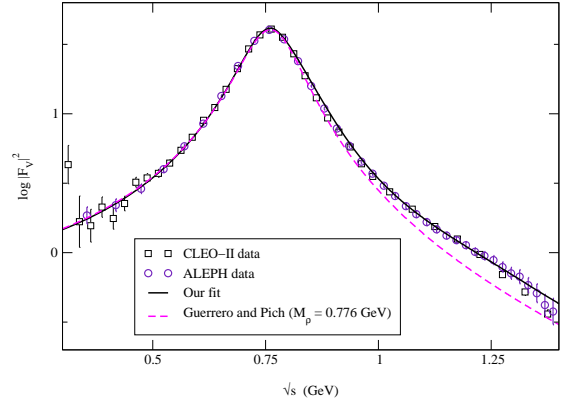}
\vspace{-.5cm} \caption{Pion form factor data \cite{ALEPHpiff}
compared with theoretical predictions \protect\cite{Portoles}.}
\label{fig:pionth}
\end{figure}

For the decay modes with lowest multiplicity,
$\tau^-\to\nu_\tau\pi^-$ and $\tau^-\to\nu_\tau K^-$, the relevant
matrix elements are already known from the measured decays
$\pi^-\to\mu^-\bar\nu_\mu$  and  $K^-\to\mu^-\bar\nu_\mu$. The
corresponding $\tau$ decay widths can then be accurately predicted.
As shown in Table~\ref{tab:ccuniv}, the predictions are in good
agreement with the measured values and provide a test of lepton
universality. Assuming universality in the quark couplings, these
decay modes determine the ratio \cite{PDG,JOP:06}
$$ \frac{|V_{us}|\, f_K}{|V_{ud}|\, f_\pi} \, = \,\left\{ \bat
0.27618\pm 0.00048 & \;\: [\Gamma_{K/\pi\to\nu_\mu\mu} ],
\\
0.267\pm 0.005 & \;\: [\Gamma_{\tau\to\nu_\tau K/\pi}]. \ea\right.
$$
The very different accuracies reflect the present poor precision on
$\Gamma(\tau^-\to\nu_\tau K^-)$.

For the two--pion final state, the hadronic matrix element is
parameterized in terms of the so-called pion form factor \ [$s\equiv
(p_{\pi^-}\! + p_{\pi^0})^2$]:
\bel{eq:Had_matrix} \langle \pi^-\pi^0| \bar d \gamma^\mu  u | 0
\rangle \equiv \sqrt{2}\, F_\pi(s)\, \left( p_{\pi^-}-
p_{\pi^0}\right)^\mu \, . \ee
A dynamical understanding of the pion form factor can be achieved
\cite{Portoles,GP:97,DPP:00,Juanjo}, using analyticity, unitarity
and some general properties of QCD, such as chiral symmetry
\cite{GL:85} and the short-distance asymptotic behavior
\cite{EGPR:89,Tempe}. Putting all these fundamental ingredients
together, one gets 
\cite{GP:97}
$$
F_\pi(s) = {M_\rho^2\over M_\rho^2 - s - i M_\rho \Gamma_\rho(s)}
\exp{\left\{-{s \,\mbox{\rm Re} A(s)           
\over 96\pi^2f_\pi^2} \right\}} ,
$$
where
$$
A(s) \equiv \log{\left({m_\pi^2\over M_\rho^2}\right)} + 8 {m_\pi^2
\over s} - {5\over 3} + \sigma_\pi^3 \log{\left({\sigma_\pi+1\over
\sigma_\pi-1}\right)}
$$
contains the one-loop chiral logarithms,
$\sigma_\pi\equiv\sqrt{1-4m_\pi^2/s}$ and the off-shell $\rho$ width
\cite{GP:97,DPP:00} is given by $\Gamma_\rho(s)\, =\,
\theta(s-4m_\pi^2)\,\sigma_\pi^3\, M_\rho\, s/(96\pi f_\pi^2)$.
This prediction, which only depends on $M_\rho$, $m_\pi$ and the
pion decay constant $f_\pi$, is compared with the data in
Fig.~\ref{fig:pionth}. The agreement is rather impressive and
extends to negative $s$ values, where the $e^-\pi$ elastic data (not
shown in the figure) sits.

\begin{figure}[t]\centering
\centering
\includegraphics[width=6.5cm,clip]{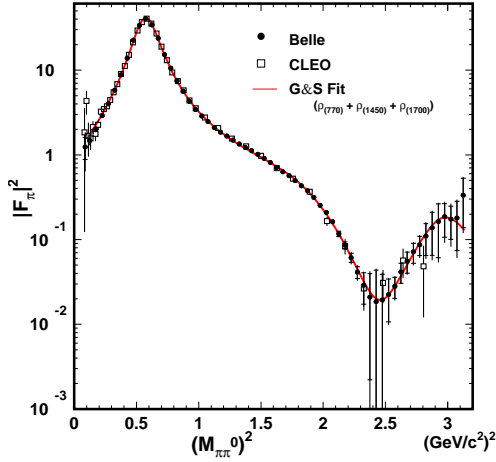}
\vspace{-.5cm} \caption{Preliminary BELLE measurement of the pion
form factor from $\tau^-\to\nu_\tau \pi^-\pi^0$
\protect\cite{Fujikawa}.} \label{fig:BellePFF}
\end{figure}

The small effect of heavier $\rho$ resonance contributions and
additional next-to-leading in $1/N_C$ corrections can be easily
included, at the price of having some free parameters which decrease
the predictive power \cite{Portoles,Juanjo}. This gives a better
description of the $\rho'$ shoulder around 1.2 GeV (continuous line
in Fig.~\ref{fig:pionth}).
A clear signal for the $\rho''(1700)$ resonance in
$\tau^-\to\nu_\tau \pi^-\pi^0$ events has been reported by BELLE,
with a data sample 20 times larger than in previous experiments
\cite{Fujikawa}.


\begin{figure}[tbh]\centering
\centering
\includegraphics[width=7.3cm,clip]{psfigs/dGTau.eps}
\vspace{-.5cm}
\caption{Predicted $\tau\to\nu_\tau K\pi$ 
distribution, together with the separate contributions from the
$K^*(892)$ and $K^*(1410)$ vector mesons as well as the scalar
component residing in 
$F_0^{K\pi}(s)$ \protect\cite{JPP:06}.} \label{fig:KpSpectrum}
\vskip .8cm 
\includegraphics[width=5.25cm,clip]{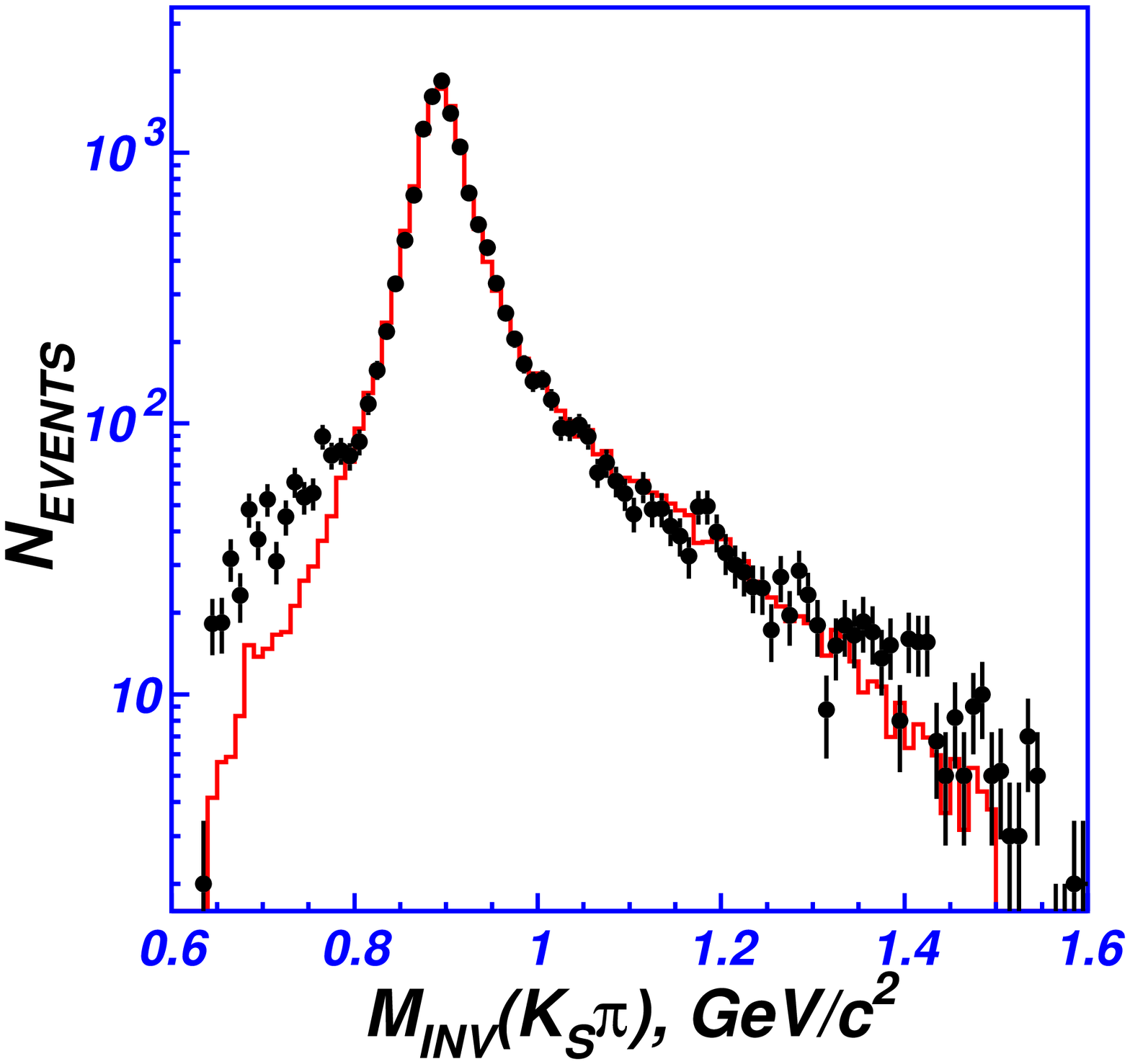}
\vspace{-.5cm}
 \caption{$K_S\pi$ invariant-mass distribution from
BELLE $\tau\to\nu_\tau K_S\pi$ events. The histogram shows the
expected $K^*(892)$ contribution \protect\cite{Shwartz}.}
\label{fig:Shwartz}
\end{figure}

The decay $\tau\to\nu_\tau K\pi$ is characterized by two form
factors with $J^{P}=1^{-}$ and $0^{+}$. The vector form factor
$F_+^{K\pi}(s)$ can be described in an analogous way to $F_\pi(s)$.
The scalar component $F_0^{K\pi}(s)$ has been recently studied
\cite{JPP:06}, taking into account additional information from
$K\pi$ scattering data through dispersion relations
\cite{JOP:06,JOP}. The decay width is dominated by the $K^*(892)$
contribution, with a predicted branching ratio Br$[\tau\to\nu_\tau
K^*] = (1.253 \pm 0.078)\%$, while the scalar component is found to
be Br$[\tau\to\nu_\tau(K\pi)_{\rm S-wave}]=(3.88\pm 0.19)\cdot
10^{-4}$. The preliminary measurements of the $\tau^-\to\nu_\tau
K_S\pi^-$ (BELLE \cite{Shwartz}) and $\tau^-\to\nu_\tau K^-\pi^0$
(BABAR \cite{Nugent}) distributions, presented at this workshop,
show a clear evidence for the scalar contribution at low invariant
mass and a $K^*(1410)$ vector component at large $s$.

The dynamical structure of other hadronic final states can be
investigated in a similar way. The $\tau\to\nu_\tau 3\pi$ decay mode
was studied in Ref.~\cite{DPP:01}, where a theoretical description
of the measured structure functions \cite{CLEO3pi,OPAL3pi,ALEPH:05}
was provided. A detailed analysis of other $\tau$ decay modes into
three final pseudoscalar mesons is in progress \cite{RPP:06}. The
more involved $\tau\to\nu_\tau 4\pi$ and $e^+e^-\to 4\pi$
transitions have also been studied \cite{EU:02}.

BABAR has presented preliminary measurements of
$\tau^-\to\nu_\tau\pi^+2\pi^-\pi^0$ and
$\tau^-\to\nu_\tau\pi^+2\pi^-\eta$ decays. The $4\pi$ distribution
is found to have a large $\omega\pi^-$ contribution, while the
$f_1(1285)\pi^-$ component is seen to be the primary source of
$\tau^-\to\nu_\tau\pi^+2\pi^-\eta$ events \cite{Sobie}. The large
statistics collected by BABAR ($2\times 10^8 \,\tau^+\tau^-$ pairs)
also allow to put limits on decay modes with 7 charged pions at the
level of few $10^{-7}$ (90\% CL)~\cite{Kass}.

CLEO has investigated rare $\tau$ decay modes with kaons in the
final state, obtaining the results $\mathrm{Br}(\tau^-\to\nu_\tau
K^-\pi^+\pi^-\pi^0) = (7.4\pm 0.8\pm 1.1)\times 10^{-4}$ ($K^0$
excluded) and $\mathrm{Br}(\tau^-\to\nu_\tau K^-K^+\pi^-\pi^0) =
(5.5\pm 1.4\pm 1.2)\times 10^{-5}$ \cite{Gan}.

\section{THE TAU HADRONIC WIDTH}
\label{sec:hadronic_width}

The inclusive character of the total $\tau$ hadronic width renders
possible an accurate calculation of the ratio
\cite{BR:88,NP:88,BNP:92,LDP:92a,QCD:94}
%
\bel{eq:r_tau_def} R_\tau \equiv { \Gamma [\tau^- \to \nu_\tau
\,\mathrm{hadrons}\, (\gamma)] \over \Gamma [\tau^- \to \nu_\tau e^-
{\bar \nu}_e (\gamma)] }\, , \ee
using analyticity constraints and the Operator Product Expansion.
One can separately compute the contributions from 
specific quark currents:
\be\label{eq:r_tau_v,a,s}
 R_\tau \, = \, R_{\tau,V} + R_{\tau,A} + R_{\tau,S}\, .
\ee
$R_{\tau,V}$ and $R_{\tau,A}$ correspond to the Cabibbo--allowed
decays through the vector and axial-vector currents, while
$R_{\tau,S}$ contains the remaining Cabibbo--suppressed
contributions.

The theoretical prediction for $R_{\tau,V+A}$ can be expressed as
\cite{BNP:92}
\be
R_{\tau,V+A} = N_C\, |V_{ud}|^2\, S_{\mathrm{EW}} \left\{ 1 +
\delta_{\mathrm{P}} + \delta_{\mathrm{NP}} \right\} ,
\ee
where $N_C=3$ and $S_{\mathrm{EW}}=1.0201\pm 0.0003$ contains the
electroweak radiative corrections \cite{MS:88}.
%
The dominant correction ($\sim 20\%$) is the perturbative QCD
contribution $\delta_{\mathrm{P}}$, which is fully known to
$O(\alpha_s^3)$ \cite{BNP:92} and includes a resummation of the most
important higher-order effects \cite{LDP:92a}.
%

%
\begin{figure}[tbh]
\label{fig:alpha_s} \centering
\includegraphics[width=7.5cm]{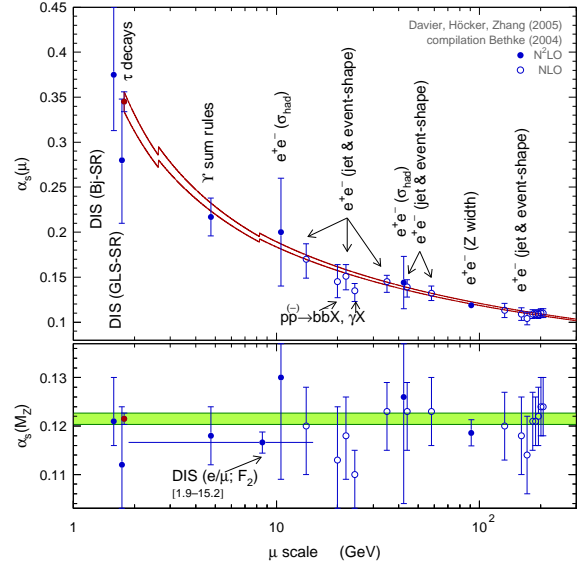}
\vspace{-0.5cm}
\caption{Measured values of $\alpha_s$ at different
scales. The curves show the energy dependence predicted by QCD,
using $\alpha_s(m_\tau^2)$ as input. The corresponding extrapolated
$\alpha_s(M_Z^2)$ values are shown at the bottom, where the shaded
band displays the $\tau$ decay result within errors \cite{DHZ:05}.}
\end{figure}

Non-perturbative contributions are suppressed by six powers of the
$\tau$ mass \cite{BNP:92} and, therefore, are very small. Their
numerical size has been determined from the invariant--mass
distribution of the final hadrons in $\tau$ decay, through the study
of weighted integrals \cite{LDP:92b},
\be R_{\tau}^{kl} \equiv \int_0^{m_\tau^2} ds\, \left(1 - {s\over
m_\tau^2}\right)^k\, \left({s\over m_\tau^2}\right)^l\, {d
R_{\tau}\over ds} \, , \ee
which can be calculated theoretically in the same way as $R_{\tau}$.
The predicted suppression \cite{BNP:92} of the non-perturbative
corrections has been confirmed by ALEPH \cite{ALEPH:05}, CLEO
\cite{CLEO:95} and OPAL \cite{OPAL:98}. The most recent analysis
\cite{ALEPH:05} gives
\bel{eq:del_np} \delta_{\mathrm{NP}} \, =\, -0.0043\pm 0.0019 \, .
\ee

The QCD prediction for $R_{\tau,V+A}$ is then completely dominated
by the perturbative contribution; non-perturbative effects being
smaller than the perturbative uncertainties from uncalculated
higher-order corrections. The result turns out to be very sensitive
to the value of $\alpha_s(m_\tau^2)$, allowing for an accurate
determination of the fundamental QCD coupling \cite{NP:88,BNP:92}.
The experimental measurement $R_{\tau,V+A}= 3.471\pm0.011$ implies
\cite{DHZ:05}
%
\be\label{eq:alpha} \alpha_s(m_\tau^2)  =  0.345\pm
0.004_{\mathrm{exp}}\pm 0.009_{\mathrm{th}} \, . \ee

The strong coupling measured at the $\tau$ mass scale is
significantly larger than the values obtained at higher energies.
From the hadronic decays of the $Z$, one gets $\alpha_s(M_Z^2) =
0.1186\pm 0.0027$ \cite{LEPEWWG}, which differs from the $\tau$
decay measurement by more than twenty standard deviations. After
evolution up to the scale $M_Z$ \cite{Rodrigo:1998zd}, the strong
coupling constant in \eqn{eq:alpha} decreases to \cite{DHZ:05}
\be\label{eq:alpha_z} \alpha_s(M_Z^2)  =  0.1215\pm 0.0012 \, , \ee
in excellent agreement with the direct measurements at the $Z$ peak
and with a similar accuracy. The comparison of these two
determinations of $\alpha_s$ in two extreme energy regimes, $m_\tau$
and $M_Z$, provides a beautiful test of the predicted running of the
QCD coupling; i.e. a very significant experimental verification of
{\it asymptotic freedom}.

\section{CABIBBO--SUPPRESSED DECAYS}
\label{sec:ms}

The separate measurement of the $|\Delta S|=0$ and $|\Delta S|=1$ \
$\tau$ decay widths allows us to pin down the SU(3) breaking effect
induced by the strange quark mass
\cite{Davier,PP:99,ChDGHPP:01,ChKP:98,MW:06,GJPPS:05,BChK:05},
through the differences \cite{PP:99}
\beqn \lefteqn{\delta R_\tau^{kl}  \equiv
  {R_{\tau,V+A}^{kl}\over |V_{ud}|^2} - {R_{\tau,S}^{kl}\over |V_{us}|^2}
  } &&
\\ & &\approx  24\, {m_s^2(m_\tau^2)\over m_\tau^2} \, \Delta_{kl}(\alpha_s)
    - 48\pi^2\, {\delta O_4\over m_\tau^4} \, Q_{kl}(\alpha_s)
\, .\no \eeqn
The perturbative QCD corrections $\Delta_{kl}(\alpha_s)$ and
$Q_{kl}(\alpha_s)$ are known to $O(\alpha_s^3)$ and $O(\alpha_s^2)$,
respectively \cite{PP:99,BChK:05}.
Since the longitudinal contribution to $\Delta_{kl}(\alpha_s)$ does
not converge well, the $J=0$ QCD expression is replaced by its
corresponding phenomenological hadronic parametrization
\cite{GJPPS:05}, which is much more precise because it is dominated
by far by the well-known kaon pole. The small non-perturbative
contribution, $\delta O_4 \equiv\langle 0| m_s \bar s s - m_d \bar d
d |0\rangle
 = -(1.5\pm 0.4)\times 10^{-3}\;\mbox{\rm GeV}^4$,
has been estimated with Chiral Perturbation Theory techniques
\cite{PP:99}.

%
%
\begin{figure}[t]
\label{fig:DS=1} \centering
\includegraphics[width=7.3cm]{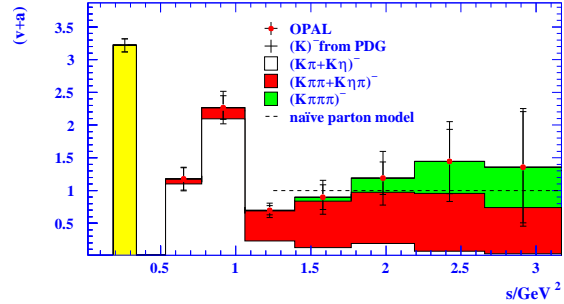}
\vspace{-0.5cm} \caption{OPAL measurement of the spectral function
distribution in $|\Delta S|=1$ $\tau$ decays \cite{OPALms}.}
\end{figure}

From the measured moments $\delta R_\tau^{k0}$ ($k=0,1,2,3,4$)
\cite{ALEPHms,OPALms}, it is possible to determine the strange quark
mass; however, the extracted value depends sensitively on the
modulus of the Cabibbo--Kobayashi--Maskawa matrix element
$|V_{us}|$. It appears then natural to turn things around and, with
an input for $m_s$ obtained from other sources, to actually
determine $|V_{us}|$ \cite{GJPPS:05}. The most sensitive moment is
$\delta R_\tau^{00}$:
\bel{eq:Vus_formula} |V_{us}|^2 =
\frac{R^{(0,0)}_{\tau,S}}{\frac{R^{(0,0)}_{\tau,V+A}}{|V_{ud}|^2}-\delta
R^{(0,0)}_{\tau,{\mathrm{th}}}} \, . \ee
Using $m_s(2~\mathrm{GeV})= (94\pm 6)~\mathrm{MeV}$, which includes
the most recent determinations of $m_s$ from lattice and QCD Sum
Rules \cite{JOP:06}, one obtains $\delta R^{00}_{\tau,{\mathrm{th}}}
= 0.240 \pm 0.032$ \cite{GJPPS:05}. This prediction is much smaller
than $R^{(0,0)}_{\tau,V+A}/|V_{ud}|^2$, making the theoretical
uncertainty in \eqn{eq:Vus_formula} negligible in comparison with
the experimental inputs $R^{(0,0)}_{\tau,V+A}=3.471\pm 0.011$ and
$R^{(0,0)}_{\tau,S}=0.1686\pm 0.0047$ \cite{DHZ:05}. Taking
$|V_{ud}|=0.97377\pm 0.00027$ \cite{PDG}, one gets~\cite{GJPPS:05}
\bel{eq:Vus_value}
 |V_{us}| =0.2220 \pm 0.0031_{\mathrm{exp}} \pm
0.0011_{\mathrm{th}}
\, . \ee
This result is competitive with the standard $K_{e3}$ determination,
$|V_{us}| =0.2236 \pm 0.0029$ \cite{JOP:06}. The precision should be
considerably improved in the near future because the error is
dominated by the experimental uncertainty, which can be reduced with
the much better data samples from BABAR \cite{Nugent}, BELLE
\cite{Shwartz,Ohshima} and the forthcoming BESIII detector
\cite{MO}. Therefore, the $\tau$ data has the potential to provide
the best determination of $|V_{us}|$.

With future high-precision $\tau$ data, a simultaneous fit of $m_s$
and $|V_{us}|$ should also become possible. A better understanding
of the perturbative QCD corrections $\Delta_{kl}(\alpha_s)$ would be
very helpful to improve the resulting $m_s$ accuracy
\cite{MW:06,GJPPS:05}.

\section{LEPTON MAGNETIC MOMENTS}
\label{sec:g-2}

The most stringent QED test comes from the high-precision
measurements of the $e$ and $\mu$ anomalous magnetic moments
$a_l\equiv (g^\gamma_l-2)/2$ \cite{HK:99,DM:04,PA:05}.
%
A recent measurement of $a_e$, using a one-electron quantum
cyclotron, has reduced the experimental uncertainty by a factor of
six \cite{OHUG:06}:
\bel{eq:a_e} a_e = (1\; 159\; 652\; 180.85\pm 0.76) \,\cdot\,
10^{-12}\, . \ee
To a measurable level, $a_e$ arises entirely from virtual electrons
and photons; these contributions are known to $O(\alpha^4)$
\cite{HK:99,DM:04,PA:05,KN:06}. The theoretical error is dominated
by the uncertainty in the input value of $\alpha$. Turning things
around, the measured value of $a_e$ provides the most precise
determination of the fine structure constant \cite{GHKNO:06}:
\be \alpha^{-1} = 137.035\; 999\; 710 \,\pm\, 0.000\; 000\; 096\, .
\ee
This number agrees with other precise determinations of $\alpha$,
but it has an uncertainty (0.70 ppb) 10 times smaller than any other
method.

The BNL-E821 experiment has recently published its final value for
$a_\mu$ 
\cite{E821:06}:
\bel{eq:a_mu} a_\mu = (11\; 659\; 208.0\pm 6.3) \,\cdot\, 10^{-10}
\, . \ee
The anomalous magnetic moment of the muon is sensitive to small
corrections from virtual heavier states; compared to $a_e$, they
scale as $m_\mu^2/m_e^2$. The Standard Model prediction can be
decomposed in five types of contributions:
\beqn 10^{10} \cdot a_\mu^{\mathrm{th}}  =
11\; 658\; 471.81 \pm 0.02 &&
\!\!\!\!\!\mathrm{QED}\:\mbox{\cite{PA:05,KN:06}}
\no\\ \mbox{} +\phantom{6}15.4\phantom{1}\pm 0.2\phantom{2} &&
\!\!\!\!\!\mathrm{EW}\:\cite{a_mu_EW}
\no\\ \mbox{} + 698.9\phantom{1} \pm 9.6\phantom{2}
&&\!\!\!\!\!\mathrm{had}^{\mathrm{ LO}}\:\cite{DHZ:05,DEHZ:03}
\no\\
\mbox{} -\phantom{69}9.8\phantom{1}\pm 0.1\phantom{2}&&
\!\!\!\!\!\mathrm{had}^{\mathrm{NLO}} \:\cite{HMNT:04}
\no\\ \mbox{} +\phantom{6}12.0\phantom{1}\pm 3.5\phantom{2} &&
\!\!\!\!\!\mathrm{lbl}\:\cite{DM:04,light_light}
\no\\  
= \, 11\; 659\; 188.3\phantom{1} \pm  10.2 && \hskip -.65cm .
\no\eeqn
This result differs by $1.6\,\sigma$ from the experimental value
\eqn{eq:a_mu}.

The main theoretical uncertainty on $a_\mu$ has a QCD origin. Since
quarks have electric charge, virtual quark-antiquark pairs induce
{\it hadronic vacuum polarization} corrections to the photon
propagator (Fig.~\ref{fig:AnMagMom}.c). Owing to the
non-perturbative character of QCD at low energies, the light-quark
contribution cannot be reliably calculated at present. Fortunately,
this effect can be extracted from the measurement of the
cross-section $\sigma(e^+e^-\to \mbox{\rm hadrons})$ and from the
invariant-mass distribution of the final hadrons in $\tau$ decays.
The largest contribution comes from the $2\pi$ final state. The
$\tau$ decay determination includes a careful investigation of
isospin breaking effects \cite{CEN:02}, using the pion form factor
expression of ref.~\cite{GP:97}, which amount to an overall
$-(2.2\pm 0.5)\% $ correction \cite{DEHZ:03}.

\begin{figure}[tb]\centering
\includegraphics[width=7.2cm]{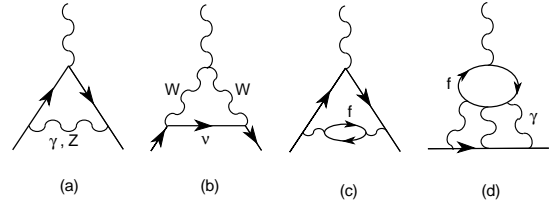}
\vspace{-.75cm} \caption{Feynman diagrams contributing to $a_l$.}
\label{fig:AnMagMom}
\end{figure}

At present, there is a discrepancy between the $2\pi$ contributions
extracted from $e^+e^-$ and $\tau$ data, which translates into
slightly different values ($2.9\,\sigma$) for
$a_\mu^{\mathrm{had,LO}}$ \cite{DEHZ:03}:
\be a_\mu^{\mathrm{had,LO}} =\left\{ \bat (690.8\pm 4.4)\cdot
10^{-10} &\quad (e^+e^-)\, ,\\  (710.3\pm 5.2)\cdot 10^{-10} & \quad
(\tau)\, . \ea\right.\ee
Therefore, from $e^+e^-$ data one gets the prediction
$a_\mu^{\mathrm{th}} = (11\, 659\, 180.2\pm 5.6)\cdot 10^{-10}$,
which disagrees with the measured value by $3.3\,\sigma$, while the
$\tau$ data gives $a_\mu^{\mathrm{th}} = (11\, 659\, 199.7\pm
6.3)\cdot 10^{-10}$, in much better agreement ($0.9\,\sigma$) with
the BNL measurement of $a_\mu$ \cite{E821:06}. In order to quote a
reference number for $a_\mu^{\mathrm{th}}$, I have used a weighted
average of these two determinations, increasing the error with the
appropriate scale factor \cite{PDG}.

New precise $e^+e^-$ and $\tau$ data sets are 
needed to settle the true value of $a_\mu^{\mathrm{had,LO}}$. The
present experimental situation is very unsatisfactory, showing
internal inconsistencies among different $e^+e^-$ and $\tau$
measurements. The KLOE $e^+e^-$ invariant-mass distribution
\cite{venanzoni} does not agree with CMD2 and SND, while the most
recent BELLE measurement of the $\tau$ decay spectrum
\cite{Fujikawa} slightly disagrees with ALEPH and CLEO
\cite{DEHZ:03}. Using CVC, one predicts from $e^+e^-$ data
$\mathrm{Br}(\tau\to\nu_\tau 2\pi) = (24.48\pm 0.18)\% $, which is
$4.5\,\sigma$ smaller than the direct $\tau$ measurement $(25.40\pm
0.10)\% $ \cite{DEHZ:03}. It is difficult to explain such a large
disagreement as an isospin-breaking effect \cite{LopezCastro}. Since
the $e^+e^-$ prediction involves a delicate integration over all the
spectrum, while the $\tau$ number is a more robust branching ratio
measurement (on which all $\tau$ experiments agree), the $e^+e^-$
analysis appears to me more suspect as a source of underestimated
systematic uncertainties.
The radiative return method \cite{German}, already used by KLOE
\cite{venanzoni}, will allow the B Factories to provide some light
on this issue. Preliminary analyses from BABAR have been already
presented at this workshop \cite{Wang}.

Additional QCD uncertainties stem from the smaller {\it
light-by-light scattering} contributions
(Fig.~\ref{fig:AnMagMom}.d). The most recent evaluations of these
corrections \cite{light_light}, have uncovered a sign mistake in
previous calculations, improving the agreement with the experimental
measurement.

If funded, the Brookhaven E969 proposal could reduce the $a_\mu$
experimental uncertainty by a factor of two or more \cite{Hertzog}.
A meaningful test of the electroweak contributions at this level of
precision requires a better control of the QCD corrections. A factor
of three improvement also seems possible in the $a_e$ measurement
\cite{OHUG:06}. On the QED side a formidable effort to perform the
fifth-order calculation has already started \cite{KN:06}.

\section{NEUTRINO OSCILLATIONS}
\label{sec:oscillations}

%
\begin{figure}[tbh]
\centering
\includegraphics[width=7.5cm]{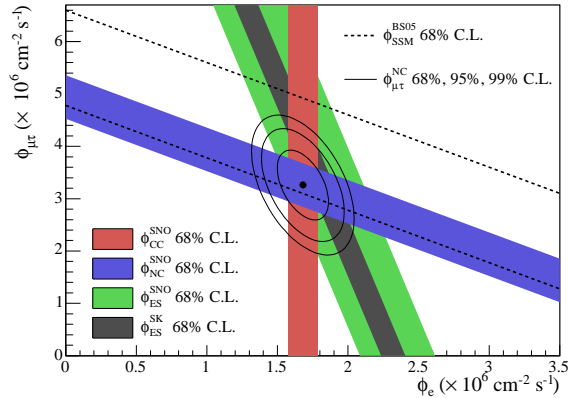}
\vspace{-0.5cm} \caption{Measured fluxes of ${}^8B$ solar neutrinos
of $\nu_\mu$ or $\nu_\tau$ type ($\phi_{\mu,\tau}$) versus the flux
of $\nu_e$ ($\phi_e$)~\cite{SNO}.} \label{fig:SNO}
\end{figure}

The flux of solar $\nu_e$ neutrinos reaching the
earth has been measured by several experiments \cite{jelley} to be
significantly below the standard solar model prediction
\cite{BP:04}.
The Sudbury Neutrino Observatory has provided strong evidence that
neutrinos do change flavour as they propagate from the core of the
Sun \cite{SNO}, independently of solar model flux predictions.
SNO is able to detect neutrinos through three different reactions:
the charged-current process $\nu_e d \to e^-pp$ which is only
sensitive to $\nu_e$, the neutral current transition $\nu_x d \to
\nu_x pn$ which has equal probability for all active neutrino
flavours, and the elastic scattering $\nu_x e^-\to\nu_x e^-$ which
is also sensitive to $\nu_\mu$ and $\nu_\tau$, although the
corresponding cross section is a factor $6.48$ smaller than the
$\nu_e$ one. The measured neutrino fluxes, shown in
Fig.~\ref{fig:SNO}, demonstrate the existence of a non-$\nu_e$
component in the solar neutrino flux at the $5.3\,\sigma$ level.
These results have been further reinforced with the KamLAND data,
showing that $\bar\nu_e$ from nuclear reactors disappear over
distances of about 180 Km \cite{KamLAND}.

Another 
evidence of oscillations has been
obtained from atmospheric neutrinos. The known discrepancy between
the experimental observations and the predicted ratio of muon to
electron neutrinos has become much stronger with the high precision
and large statistics of Super-Kamiokande  \cite{SKatm,suzuki}. The
atmospheric anomaly appears       
to originate in a
reduction of the $\nu_\mu$ flux, and the data strongly favours the
$\nu_\mu\to\nu_\tau$ hypothesis. This result has been confirmed by
K2K \cite{K2K} and MINOS \cite{MINOS}, observing the disappearance
of accelerator $\nu_\mu$'s at distances of 250 and 735 Km,
respectively. Super-Kamiokande has recently reported statistical
evidence of $\nu_\tau$ appearance at the $2.4\,\sigma$ level
\cite{SKatm}. The direct detection of the produced $\nu_\tau$ is the
main goal of the ongoing CERN to Gran Sasso neutrino program
\cite{CGSNP}.

\begin{figure}[tbh]
\centering
\includegraphics[height=6cm,clip]{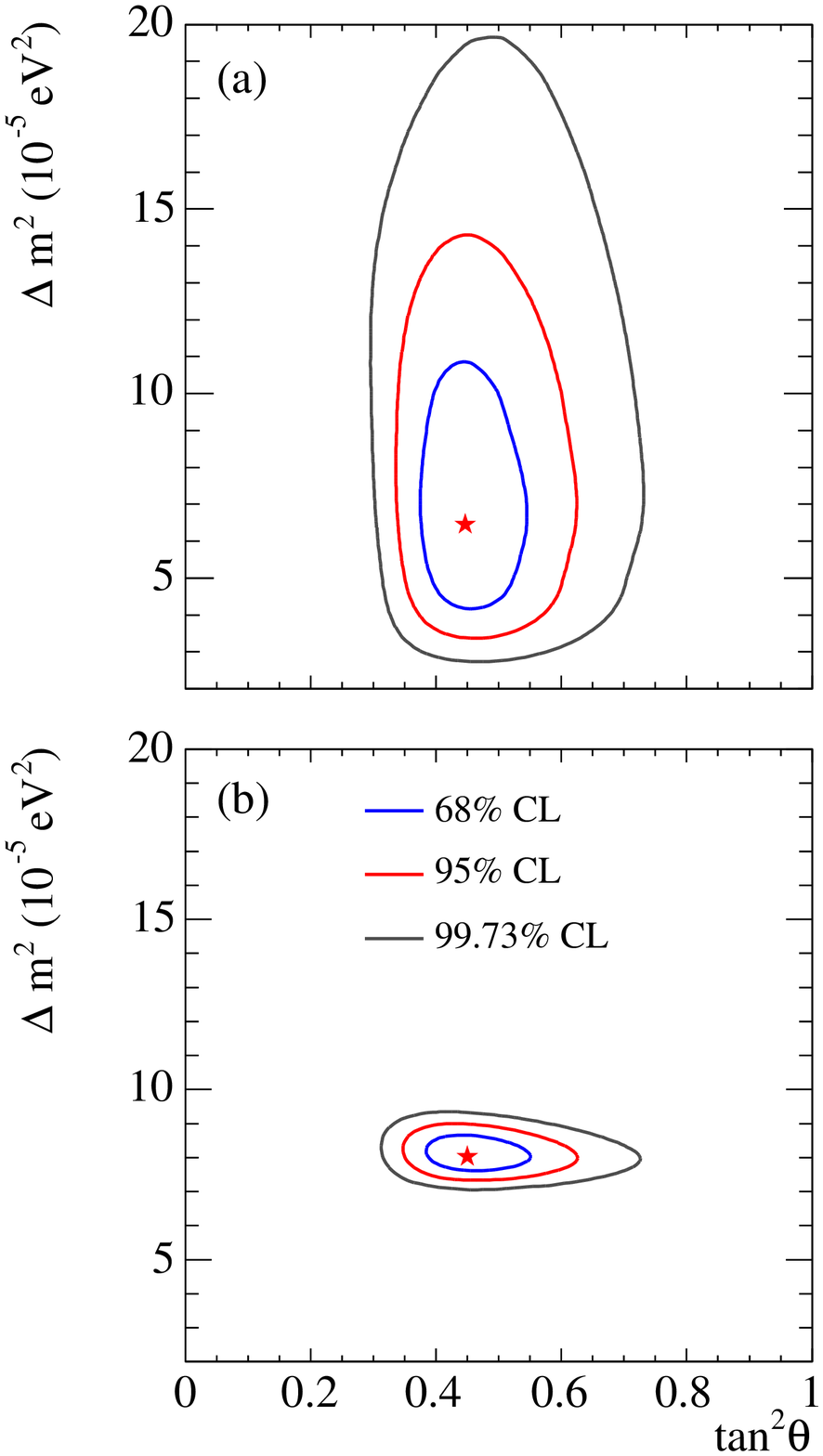}
\vskip -.5cm \caption{Allowed regions for $2\nu$ oscillations for
the combination of solar ($\nu_e$) and KamLAND  ($\bar\nu_e$) data,
assuming $\mathcal{CPT}$ symmetry \cite{SNO}.} \label{fig:SolarNu}
\end{figure}
\begin{figure}[tbh]
\centering\includegraphics[height=6cm]{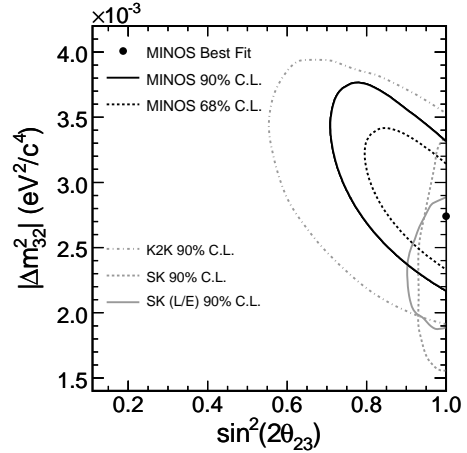}
 \vskip -.5cm
 \caption{MINOS allowed regions for $\nu_\mu$ disappearance
oscillations, compared with K2K and Super-Kamiokande results
\cite{MINOS}.} \label{fig:AtmosNu}
\end{figure}

Thus, we have now clear experimental evidence that neutrinos are
massive particles and there is mixing in the lepton sector.
The present solar, atmospheric, accelerator and reactor neutrino
data, leads to the following preferred ranges for the oscillation
parameters \cite{PDG}:
\beqn\label{nu_mix}
 \Delta m^2_{21}\; = \; \left( 8.0\, {}^{+\, 0.4}_{-\, 0.3}\right)
 \cdot 10^{-5}\;\mathrm{eV}^2\; , \hskip 1.8cm &&
 \no\\
 1.9 \cdot 10^{-3}\; <\; |\Delta m^2_{32}|\; /\; \mathrm{eV}^2 \; <\;
 3.0 \cdot 10^{-3} \; ,\hskip 0.3cm  &&
 \no\\ \sin^2{(2\theta_{12})}\; =\; 0.86\, {}^{+\, 0.03}_{-\, 0.04}
 \; , \hskip 2.9cm &&
 \no\\ \sin^2{(2\theta_{23})}\; >\; 0.92 \; , \hskip 3.75cm &&
 \no\\
 \sin^2{(2\theta_{13})}\; <\; 0.19 \; , \hskip 3.75cm &&
\eeqn
where $\Delta m^2_{ij}\equiv m^2_i - m^2_j$ are the mass squared
differences between the neutrino mass eigenstates $\nu_{i,j}$ and
$\theta_{ij}$ the corresponding mixing angles in the standard
three-flavour parametrization. The ranges indicate 90\% CL.
In the limit $\theta_{13}=0$, solar and atmospheric neutrino
oscillations decouple because $\Delta m^2_\odot \ll\Delta
m^2_\mathrm{atm}$. Thus, $\Delta m^2_{21}$, $\theta_{12}$ and
$\theta_{13}$ are constrained by solar data, while atmospheric
experiments constrain $\Delta m^2_{32}$, $\theta_{23}$ and
$\theta_{13}$. The angle $\theta_{13}$ is strongly constrained by
the CHOOZ reactor experiment \cite{CHOOZ}. New planned reactor
experiments \cite{jelley}, T2K and NO$\nu$A \cite{suzuki} are
expected to achieve sensitivities around $\sin^2{(2\theta_{13})}\sim
0.01$.

\section{NEW PHYSICS}

Non-zero neutrino masses constitute a clear indication of new
physics beyond the Standard Model. Right-handed neutrinos are an
obvious possibility to incorporate Dirac neutrino masses.
However, the $\nu_{iR}$ fields would be $SU(3)_C\otimes
SU(2)_L\otimes U(1)_Y$ singlets, without any Standard Model
interaction.
Moreover, the gauge symmetry would allow for a right-handed Majorana
neutrino mass term of arbitrary size, not related to the ordinary
Higgs mechanism.

\begin{table*}[tbh]
\caption{Best published limits (90\% CL) on lepton-flavour-violating
decays \cite{PDG,LFVbabar,LFVbelle}.}
\label{table:LFV}\vspace{0.2cm}
\renewcommand{\tabcolsep}{1.1pc} 
\renewcommand{\arraystretch}{1.2} 
\begin{tabular}{@{}llllll}
\hline
 $\Br (\mu^-\to e^-\gamma) < 1.2\cdot 10^{-11}$ &
 $\Br (\mu^-\to e^-2\gamma) < 7.2\cdot 10^{-11}$ &
 $\Br (\mu^-\to e^-e^-e^+) < 1.0\cdot 10^{-12}\hskip -.2cm $
 \\
 $\Br (\tau^-\to \mu^-\gamma) < 6.8\cdot 10^{-8}$ &
 $\Br (\tau^-\to e^-\gamma) < 1.1\cdot 10^{-7}$ &
 $\Br (\tau^-\to e^-e^-\mu^+) < 1.1\cdot 10^{-7}\hskip -.2cm $
 \\
 $\Br (\tau^-\to e^-K_S) < 5.6\cdot 10^{-8}$ &
 $\Br (\tau^-\to \mu^-K_S) < 4.9\cdot 10^{-8}$ &
 $\Br (\tau^-\to \mu^+\pi^-\pi^-) < 0.7\cdot 10^{-7}\hskip -.2cm $
 \\
 $\Br (\tau^-\to \Lambda\pi^-) < 7.2\cdot 10^{-8}$ &
 $\Br (\tau^-\to \mu^-e^+\mu^-) < 1.3\cdot 10^{-7}\hskip -.1cm $ &
 $\Br (\tau^-\to e^-\pi^+\pi^-) < 1.2\cdot 10^{-7}\hskip -.2cm $
 \\
 $\Br (\tau^-\to \mu^-\pi^0) < 1.1\cdot 10^{-7}$ &
 $\Br (\tau^-\to \mu^-\eta) < 1.3\cdot 10^{-7}$ &
 $\Br (\tau^-\to e^-\pi^0) < 1.4\cdot 10^{-7}$
 \\ \hline
\end{tabular}\end{table*}

Adopting a more general effective field theory language, without any
assumption about the existence of right-handed neutrinos or any
other new particles, one can write the most general $SU(3)_C\otimes
SU(2)_L\otimes U(1)_Y$ invariant lagrangian, in terms of the known
low-energy fields ($\nu_{iL}$ only).
The Standard Model is the unique answer with $d=4$. The first
contributions from new physics appear at $d=5$ and have also a
unique form \cite{WE:79}, which violates lepton number by two units:
\bel{eq:WE} \Delta \cL\; =\; - {c_{ij}\over\Lambda}\; \bar
L_i\,\tilde\phi\, \tilde\phi^t\, L_j^c \; + \; \mathrm{h.c.}\, , \ee
where $\phi$ and $L_i$ are the scalar and $i$-flavoured lepton
$SU(2)_L$ doublets, $\tilde\phi \equiv i\,\tau_2\,\phi^*$ and $L_i^c
\equiv \mathcal{C} \bar L_i^t$. Similar operators with quark fields
are forbidden, due to their different hypercharges, while
higher-dimension operators would be suppressed by higher powers of
the new-physics scale $\Lambda$.
After spontaneous symmetry breaking, $\langle\phi^{(0)}\rangle =
v/\sqrt{2}$, $\Delta \cL$
generates a Majorana mass term:
\bel{eq:Majorana} \cL_M = -{1\over 2}\, \bar\nu_{iL} M_{ij}\,
\nu_{jL}^c
 +\mathrm{h.c.} ,
\quad M_{ij} = {c_{ij}\, v^2\over\Lambda} . \ee
Thus, Majorana neutrino masses should be expected on general
symmetry grounds. The relation~(\ref{eq:Majorana}) generalizes the
well-known see-saw mechanism ($m_{\nu_{L}}\sim m^2/\Lambda$)
\cite{RGMY:79}. Taking $m_\nu\gsim 0.05$~eV, as suggested by
atmospheric neutrino data, one gets $\Lambda/c_{ij}\lsim
10^{15}$~GeV, amazingly close to the expected scale of Gran
Unification.

With non-zero neutrino masses, the leptonic charged-current
interactions involve a flavour mixing matrix $\mathbf{U}$. The
present data on neutrino oscillations imply that all elements of
$\mathbf{U}$ are large, except for $\mathbf{U}_{e3}< 0.18$
\cite{Giunti}. Therefore, the mixing among leptons appears to be
very different from the one in the quark sector.

The smallness of neutrino masses implies a strong suppression of
neutrinoless lepton-flavour-violating processes, which can be
avoided in models with other sources of lepton flavour violation,
not related to $m_{\nu_i}$ \cite{Masiero}. The
B Factories are pushing the experimental
limits on neutrinoless 
$\tau$ decays beyond the $10^{-7}$ level \cite{LFVbabar,LFVbelle},
increasing in a drastic way the sensitivity to new physics scales.
Future experiments could push further some limits to the $10^{-9}$
level \cite{Roney,Schoening}, allowing to explore interesting and
totally unknown phenomena. Complementary information will be
provided by the MEG experiment, which will search for $\mu^+\to
e^+\gamma$ events with a sensitivity of $10^{-13}$ \cite{Mori}.
There are also ongoing projects at J-PARC aiming to study $\mu\to e$
conversions in muonic atoms, at the $10^{-18}$ level \cite{Sato}.

An important question to be addressed in the future concerns the
possibility of leptonic CP violation and its relevance for
explaining the baryon asymmetry of our universe through
leptogenesis. 


\section{OUTLOOK}

Our knowledge of the lepton properties has been considerably
improved during the last few years. Lepton universality has been
tested to a rather good accuracy, both in the charged and neutral
current sectors. The Lorentz structure of the leptonic $l\to\nu_l
l'\bar\nu_{l'}$ decays has been determined with good precision in
the $\mu$ decay and relevant constraints have been obtained for the
$\tau$ \cite{PDG}. An upper limit of 3.2\% (90\% CL) has been
already set on the probability of having a (wrong) decay from a
right-handed $\tau$ \cite{taurev98,taurev03,Stahl00}.

The quality of the hadronic $\tau$ decay data has made possible to
perform quantitative QCD tests and determine the strong coupling
constant very accurately, providing a nice experimental verification
of asymptotic freedom. Information on the strange quark mass has
also been obtained from Cabibbo-suppressed hadronic $\tau$ decays;
these decay modes are expected to provide soon the most precise
determination of $|V_{us}|$.

The recent measurement of the electron anomalous magnetic moment has
substantially improved the determination of $\alpha$,
while the BNL investigation of $a_\mu$
has reached the needed sensitivity to explore higher-order
electroweak corrections. Further experimental progress could be
possible.
To perform a meaningful precision test of the
electroweak theory, it is necessary to control better the QCD
contributions. The present $e^+e^-$ versus $\tau$ experimental
controversy on the photon vacuum polarization should be resolved,
and a more accurate determination of the light-by-light scattering
contribution is needed.

The first hints of new physics beyond the Standard Model have
emerged recently, with convincing evidence of neutrino oscillations
from solar, atmospheric, accelerator and reactor neutrino
experiments. The existence of lepton flavour violation opens a very
interesting window to unknown phenomena, which we are just starting
to explore. It seems possible to push the present limits on
neutrinoless $\tau$ decays beyond the $10^{-8}$ or even $10^{-9}$
level, probing the underlying lepton flavour dynamics to a much
deeper level of sensitivity. At the same time, new neutrino
oscillation experiments will measure the small mixing angle
$\theta_{13}$ and will investigate whether CP violating phases are
also present in the lepton mixing matrix.

The huge $\tau$ data sample accumulated at the B Factories will soon
be complemented with the BESIII $\tau^+\tau^-$ pairs, collected at
threshold. Moreover, a possible Super-B Factory is already under
study \cite{Roney}, and further ideas towards a low-energy Tau-Charm
Factory with luminosities beyond $10^{35}\;\mathrm{cm^{-2}\,
s^{-1}}$, using a high-intensity $e^+e^-$ linear collider, have been
presented at this workshop \cite{Schoening}. Therefore, $\tau$
physics will continue being a very active field of research in the
next years. 
Among the new topics which could be investigated in the future, it
is worth mentioning the search for CP violating signals in $\tau$
decays.

Decays of heavier particles into $\tau$ leptons are another
interesting field of research, as exemplified by the recent
determination of $f_B |V_{ub}|$ from the $B\to\tau\nu_\tau$
branching ratio \cite{BtoTau}.
The Tevatron has also shown the advantages of the $\tau$ lepton as a
tool for new physics searches \cite{Hays}. The $\tau$ provides a
clean signature and the possibility to perform polarization
analyses, which makes $\tau$ identification a key ingredient for new
discoveries. Strategies to exploit the full $\tau$ potential at LHC
are being actively investigated at present \cite{LHC}.



\section*{ACKNOWLEDGEMENTS}
I want to thank Alberto Lusiani for organizing an enjoyable
conference. I have benefited from 
discussions with M.~Davier, M.~Jamin, J.~Portol\'es and G.~Rodrigo. 
This work has been supported by
MEC, Spain (Grant FPA2004-00996) 
and by the EU MRTN-CT-2006-035482 (FLAVIA{\it net}).


\end{document}